\documentclass[doublespace]{article}

\usepackage{amssymb}
\usepackage{amsmath}   
\usepackage{amsfonts}
\usepackage{amsthm}
 \usepackage{epsfig} 
 \usepackage{graphics} 

\newtheorem{proposition}{Proposition}

\theoremstyle{definition}

\newtheorem{remark}{Remark}

\title{New Linear Codes from Matrix-Product Codes with Polynomial Units}
\date{}

\usepackage{url}
\newcommand{\fq}{\mathbb{F}_q}
\newcommand{\talque}{~ | ~}

\newcommand{\code}{\mathcal{C}}

\begin{document}
\maketitle

\centerline{\scshape Fernando Hernando\footnote{Is supported in part by the Claude Shannon Institute, Science Foundation Ireland Grant 06/MI/006 (Ireland) and by MEC MTM2007-64704 and Junta de CyL VA025A07 (Spain).}}
\medskip
{\footnotesize
 \centerline{Department of Mathematics, University College Cork}
   \centerline{Cork, Ireland}
} 

\medskip

\centerline{\scshape Diego Ruano\footnote{Is supported in part by MEC MTM2007-64704 and Junta de CyL VA065A07 (Spain).}}

\medskip
{\footnotesize
 \centerline{Department of Mathematical Sciences, Aalborg University}
   \centerline{9220-Aalborg {\O}st, Denmark}
}


\medskip

\begin{abstract}
A new construction of codes from old ones is considered, it is an extension of the matrix-product construction. Several linear codes that improve the parameters of the known ones are presented.
\end{abstract}

\section{Introduction}

Matrix-Product codes were initially considered in \cite{Blackmore-Norton,Ozbudak}. They are an extension of several classic constructions of codes from old ones, like the Plotkin $u|u+v$-construction. In this article we consider this construction with cyclic codes, matrix-product codes with polynomials units, where the elements of the matrix used to define the codes are polynomials instead of elements of the finite field. The codes obtained with this construction are quasi-cyclic codes \cite{LF}. These codes became important after it was shown that some codes in this class meet a modified Gilbert-Varshamov bound \cite{Kas}.

An extension of the lower bound on the minimum distance from \cite{Ozbudak} is obtained. This bound is sharp for a matrix-product code of nested codes, however it is not sharp in this new setting, that is we obtain codes with minimum distance beyond this bound. By investigating the construction of the words with possible minimum weight of a matrix-product code, we are able to sift an exhaustive search and to obtain three matrix-product codes with polynomials units, that improve the parameters of the codes in \cite{cota}. Another four linear codes, improving the parameters of the known linear codes, are obtained from the previous ones.

\section{Matrix-Product Codes with Polynomial Units}\label{sec:mp}

Let $\fq$ be the finite field with $q$ elements, $C_1, \ldots, C_s \subset \mathbb{F}_q^m$ cyclic codes of length
$m$ and  $A=(a_{i,j})$ an $s\times l$-matrix, with $s\leq l$, whose entries are units in $\fq[x]/(x^m -1)$. A unit in $\fq[x]/(x^m -1)$ is a polynomial of degree lower than $m$ whose greatest common divisor with $x^m -1$ is $1$ (they are co-prime). We remark, that the cyclic codes generated by $f$ and by $f  u$, with $f \mid x^m -1$ and $\gcd (u,x^m-1)=1$, are the same code. The so-called matrix-product code with polynomial units $C=[C_1 \cdots C_s] \cdot A$ is the set of all matrix-products $[c_1 \cdots c_s] \cdot A$ where $c_i\in C_i \subset  \fq[x]/(x^m -1)$  for $i=1,\ldots, s$.

The $i$-th column of any codeword is an element of the form $\sum_{j=1}^s a_{j,i} c_j\in \mathbb{F}_q[x]/(x^m-1)$, the codewords can be viewed as,
\begin{equation*}\label{VectorCodeword}
c=\left(\sum_{j=1}^s a_{j,1} c_j, \ldots , \sum_{j=1}^s a_{j,l} c_j \right)
\in\mathbb({F}_q[x]/(x^m -1))^l.
\end{equation*}

One can generate $C$ with the matrix:
\begin{displaymath}
G=\left(
\begin{tabular}{cccccc}
$a_{1,1}f_1$ & $a_{1,2}f_1$ & $\cdots$ & $a_{1,s}f_1$& $\cdots$ & $a_{1,l}f_1$\\
$a_{2,1}f_2$ & $a_{2,2}f_2$& $\cdots$ & $a_{2,s}f_2$ & $\cdots$ & $a_{2,l}f_2$\\
$\vdots$ & $\vdots$& $\cdots$ & $\vdots$& $\cdots$ & $\vdots$\\
$a_{s,1}f_s$ & $a_{s,2}f_s$& $\cdots$ & $a_{s,s}f_s$ & $\cdots$ & $a_{s,l}f_s$\\
\end{tabular}\right),
\end{displaymath}
where $f_i$ is the generator polynomial of $C_i$, $i=1,\ldots,s$. That is, we have that $C = \{ (h_1, \ldots, h_s) G ~|~ h_i \in \fq [x] \mathrm{~with~degree~} < m - deg(f_i), i=1,\ldots ,s  \}$ and it follows that $C$ is a quasi-cyclic code.

\begin{proposition}
Let $C_i$ be a $[m,k_i,d_i]$ cyclic code, then the matrix-product code with polynomial units $C=[C_1 \cdots C_s] \cdot A$ is a linear code over $\mathbb{F}_q$ with length $lm$ and dimension $k=k_1+\cdots+k_s$  if the matrix $A$ has
full rank over $\mathbb{F}_q[x]/(x^m-1)$.
\end{proposition}
\begin{proof}
The length follows from the construction of the code. Let $A$ be a $s\times l$ matrix with $s\leq l$ which has full rank. Let $c_i\in C_i$ for $i=1,\ldots,s$ such that $[c_1,\ldots,c_s]\neq [0,\ldots,0]$. Since $A$ has rank equal to $s$ then $[c_1,\ldots,c_s]\cdot A\neq [0,\ldots,0]$. Therefore, $\# C=\#\{[c_1,\ldots,c_s]\cdot A\mid c_i\in C_i, i=1,\ldots,s\}=  (\# C_1) \cdots (\# C_s)=q^{k_1+\cdots+k_s}$.
\end{proof}

We denote by $R_i= (a_{i,1},\ldots,a_{i,l})$ the element of $(\fq [x]/(x^m -1))^l$ consisting of the $i$-th row of $A$, for $i=1,\ldots,s$. We consider $C_{R_i}$, the $\fq [x]/(x^m -1)$-submodule of $(\fq [x]/(x^m -1))^l$ generated by $R_1,\ldots, R_i$. In other words, $C_{R_i}$ is a linear code over a ring, and we denote by $D_i$ the minimum Hamming weight of the words of $C_{R_i}$, $D_i = \min \{ wt (x) \talque x \in C_{R_i}  \}$.
We obtain a lower bound for the minimum distance of $C$ by just extending the proof of the main result in \cite{Ozbudak}.

\begin{proposition}\label{lowerbound}
Let $C$ be the matrix-product code with polynomial units $[C_1 \cdots C_s] \cdot A$ where $A$ has
full rank over $\mathbb{F}_q[x]/(x^m-1)$. Then
\begin{equation}\label{distancia}
d(C)\geq d^\ast= \min\{d_1D_1,d_2D_2, \ldots ,d_s D_s\},
\end{equation}
where $d_i = d(C_i)$, $D_i = d(C_{R_i})$ and $C_{R_i}$ is  as described above.
\end{proposition}
\begin{proof}
Any codeword of $C$ is of the form $c=[c_1 \cdots
c_s]\cdot A$. Let us suppose that $c_r \neq 0$ and $c_i=0$ for all
$i>r$. It follows that $[c_{j,1}x^{j-1}, \cdots,  c_{j,s}x^{j-1}] \cdot A \in
C_{R_r}$ for $j=1,\ldots,m$, where $c_i = c_{1,i} + c_{2,i} x + \cdots + c_{m,i} x^ {m-1}$. Since $c_r \neq 0$ it has at least
$d_r$ monomials with non-zero coefficient. Suppose $c_{{i_v},r} \neq 0$, for
$v=1,\ldots, d_r$. For each $v=1,\ldots, d_r$, the product
$[c_{{i_v},1}x^{j-1}, \cdots, c_{{i_v},s}x^{j-1}]\cdot A$ is a non-zero codeword
in $C_{R_r}$, since $A$ has full rank. Therefore the weight of
$[c_{{i_v},1}x^{i_v-1}, \cdots, c_{{i_v},s}x^{i_v-1}]\cdot A$ is greater than or equal
to $D_r$ and the weight of $c$ is greater than or equal to
$d_rD_r$.
\end{proof}

\begin{remark}
If $C_1, \ldots, C_s \subset \mathbb{F}_q^m$ are linear codes of length $m$ and  $A=(a_{i,j}) \in \mathcal{M}(\fq, s \times l)$ a matrix with
$s\leq l$, then $C=[C_1 \cdots C_s] \cdot A$ is a matrix-product code, initially considered in \cite{Blackmore-Norton,Ozbudak}. We denote by $R_i= (a_{i,1},\ldots,a_{i,l})$ the element of $\mathbb{F}_q^l$ consisting of the $i$-th row of $A$, for $i=1,\ldots,s$. We set $D_i$ the minimum distance of the code $C_{R_i}$ generated by $\langle R_1,\ldots, R_i\rangle$ in $\fq^l$. In \cite{Ozbudak} the following lower bound for the minimum distance of the matrix-product code $C$ is obtained, $d(C)\geq \min\{d_1D_1,d_2D_2, \ldots ,d_s D_s\}$, where $d_i$ is the minimum distance of $C_i$. If we consider $C_1, \ldots, C_s$ nested codes, the previous bound is sharp for matrix-product codes \cite{hlr}. However, if we consider a matrix-product code with polynomial units, then the bound from proposition \ref{lowerbound} is not sharp in general, as one can see in the examples stated below.
\end{remark}

Let us consider the same approach as that of \cite{hlr} to construct a codeword with minimum weight in this more general setting: set $c_{1}, \ldots, c_{p}\in \fq[x]/(x^m -1)$ such that $c_{1}=\cdots = c_{p}$, with $wt(c_{p}) = d_p$, and $c_{p+1}=\ldots=c_s = 0$. Let $r = \sum_{i=1}^p r_i R_i$, with $r_i \in \fq[x]/(x^m-1)$, be a word in $C_{R_{p}}$ with weight $D_p$. If $c'_i = r_i c_i$ then
$$[c'_1 \cdots c'_s]\cdot A=c_1\left(\sum_{j=1}^p a_{j,1} r_j,
\ldots,\sum_{j=1}^p a_{j,l} r_j \right)=c_p r.$$

Although, for a cyclic code $C$ and a unit $g$ in $\fq[x]/(x^m-1)$, $C = \{ c  g \talque c \in C \}$, the weight of $c$ is different from the one of $c g$, in general.  Hence, the weight of $c_p r$ is greater than or equal to $d_p D_p$. We remark that this phenomenon allows us to obtain codes with minimum distance beyond the lower bound.

\section{New linear codes: Plotkin construction with polynomials}\label{sec:newcodes}

Obtaining a sharper bound than the one in the previous section is a very tough problem, actually it is the same question as the computation of the minimum distance of a quasi-cyclic code. However, by analyzing the lower bound $d^\ast$ we have performed a search to find codes with good parameters. An exhaustive search in this family is only feasible if one considers some extra conditions, these conditions should be necessary for having good parameters, but not sufficient. We will assume further particular conditions that allowed us to successfully achieve a search, discarding a significant amount of cases. We have used the structure obtained in the previous section for matrix-product codes with polynomials units from nested codes and we have obtained some binary linear codes improving the parameters of the  previously known codes.

Let $s=l=2$, and $A$ the matrix
$$
A=\left(\begin{matrix}
g_1 & g_2  \\
0 & g_4 \\
\end{matrix}\right),
$$
where $g_1, g_2, g_4$ are units in $\mathbb{F}_2[x]/(x^m-1)$. In this way $A$ is full rank over $\mathbb{F}_2[x]/(x^m-1)$ with $D_1=2$ and $D_2=1$. We may also consider this family of codes as an extension of the Plotkin $u\mid u+v$-construction.

For nested matrix-product codes the bound $d^\ast = \min \{ d_1 D_1, \ldots, d_s D_s \}$ is sharp. Furthermore, by \cite[Theorem 1]{hlr} we have some words with weight $d_i D_i$ for $i=1, \ldots, s$. We follow the construction of these words and consider a matrix $A$ in a such a way that they have weight larger than $d_iD_i$. Let $C_1 = (f_1)$ and $C_2 = (f_2)$, with $f_1 \mid f_2$ (that is, $C_1 \supset C_2$), and $C=[C_1C_2]\cdot A$. We consider $h_1, h_2 \in \mathbb{F}_2[x]$ such that $wt(f_1 h_1) = d_1$ and $wt (f_2 h_2) =d_2$, and $r_1, r_2 \in \mathbb{F}_2[x]/(x^m -1)$ such that $r_1 R_1 + r_2 R_2$ is a codeword with minimum Hamming weight in $C_{R_2}$, that is with weight $1$. Thus, the words
$[f_1 h_1, 0] \cdot A= (f_1 h_1 g_1, f_1 h_1 g_2)$ and $[f_2 h_2 r_1, f_2 h_2 r_2]\cdot A= ( f_2 h_2 r_1 g_1, f_2 h_2 (r_1 g_2 + r_2 g_4) )$ have weight greater than or equal to $2d_1$ and $d_2$, respectively.

In particular, the words with minimum Hamming weight in $C_{R_2}$ are generated by $R_2$, for $r_1=0$,  and $g_4 R_1 -g_2 R_2$, for $r_1=g_4$, $r_2=-g_2$. Therefore, the words of $C$ with possible minimum weight are: $(f_1 h_1 g_1 , f_1 h_1 g_2)$, $(0, f_2 h_2 g_4)$ and $(f_2 h_2 g_1 g_4, 0)$. Hence, we want to get $f_1 h_1 g_1$ or $f_1 h_1 g_2$ with weight greater than $d_1$  and $f_2 h_2 g_4 $ and $f_2 h_2 g_1 g_4$ with weight greater than $d_2$.

We shall assume also that $d_2 > 2 d_1$, therefore we only should have $f_1 h_1 g_1$ or $f_1 h_1 g_2$ with weight greater than $d_1$ in order to have a chance to improve the lower bound from Proposition \ref{lowerbound}.

Moreover we may consider $g_1=1$ without further restriction of generality: notice that $f_2$ and $f_2g_1$ define the same cyclic code, hence a codeword is of the form $(f_1h_1g_1,f_1h_1g_2+f_2h_2g_1)$. Multiplying by $g_1^{-1}$ we obtain $(f_1h_1,f_1h_1(g_2/g_1)+f_2h_2)$ where $g=g_2/g_1$ is a unit.

Summarizing, we have performed a sifted search following the criteria: we consider matrix-product codes with polynomial units $C=[C_1C_2]\cdot A$, where $C_1, C_2$ are cyclic nested codes, with same length and $d_2$ larger than $2 d_1$, and a matrix
$$
A=\left(\begin{matrix}
1 & g \\
0 & 1
\end{matrix}\right),
$$with $g$ unit in $\mathbb{F}_2[x]/(x^m -1)$ such that $wt(f_1 h_1 g) > d_1$.

We have compared the minimum distance of these binary linear codes with the ones in \cite{cota} using \cite{ma}. We pre-computed a table containing all the cyclic codes up to length $55$, their parameters and their words of minimum weight. We obtained the following linear codes whose parameters are better than the ones previously known:

\vspace{0.2cm}

\begin{tabular}{|l|l|}
\hline From \cite{cota} & New codes   \\ \hline
$[94,25,26]$ & $\code_1=[94,25,27]$ \\ \hline
$[102,28,27]$ & $\code_2=[102,28,28] $ \\ \hline
$[102,29,26]$ & $\code_3=[102, 29, 28] $\\ \hline
\end{tabular}

\vspace{0.2cm}

$\code_1=[C_1,C_2] \cdot A$, where $C_1=(f_1)$ and $C_2=(f_2)$ with:

\begin{itemize}

\item $f_1=x^{23} + x^{22} + x^{21} + x^{20} + x^{18} + x^{17} + x^{16} +
x^{14} + x^{13} + x^{11} + x^{10}+ x^9 + x^5 + x^4 + 1,$

\item $f_2=(x^{47}-1)/(x+1),$

\item $g=x^{20} + x^{19} + x^{13} + x^{12} + x^{11} + x^9 + x^7 +
x^4 + x^3 + x^2 + 1.$

\end{itemize}

$\code_2=[C_1,C_2]\cdot A$, where $C_1=(f_1)$ and $C_2=(f_2)$ with:

\begin{itemize}
\item $f_1=x^{25} + x^{23} + x^{22} + x^{21} + x^{20} + x^{18} + x^{16} +
x^{11 }+ x^{10} + x^8 + x^7 + x^6 +x^5 + x^4 + x + 1,$

\item $f_2=(x^{51}-1)/(x^2+x+1),$

\item $g=x^{20} + x^{15 }+ x^{14} + x^{10} + x^9 + x^7 + 1.$
\end{itemize}

$\code_3=[C_1,C_2] \cdot A$, where $C_1=(f_1)$ and $C_2=(f_2)$ with:

\begin{itemize}
\item $f_1=x^{24} + x^{23} + x^{21} + x^{19} + x^{18} + x^{15} + x^{14} +
x^{13} + x^{12} + x^{11} + x^9+ x^8+ x^6 + x^4 + 1,$

\item $f_2=(x^{51}-1)/(x^2+x+1),$

\item $g=x^{50} + x^{49} + x^{48} + x^{46} + x^{44} + x^{43} + x^{42} + x^{41} + x^{38} + x^{37} + x^{36} +
    x^{34} + x^{32} + x^{29} + x^{27} + x^{25} + x^{24} + x^{19} + x^{17} + x^{15} + x^{13} + x^{12} +
    x^{10} + x^8 + x^5 + x + 1.$

\end{itemize}

Moreover operating on $C_3$ we get four more codes.

\vspace{.2 cm}

\begin{tabular}{|c|l|l|}
\hline From \cite{cota} & New codes & Method   \\ \hline
$[101,29,26]$ & $\code_4 =[101,29,27]$ &  Puncture~Code($\code_3$,{102}) \\ \hline $[101,28,26]$ & $\code_5 =[101,28,28]$ &  Shorten~Code($\code_3$,{101}) \\ \hline $[100,28,26]$ & $\code_6 =[100,28,27]$ &  Puncture~Code($\code_5$,{101}) \\ \hline
$[103,29,27]$ & $\code_7 =[103,29,28]$ &  Extend~Code($\code_3$) \\ \hline
\end{tabular}

\vspace{.2 cm}

Also, a good number of new quasi-cyclic codes reaching the best known lower bounds are achieved with this method.  One can find $434$ of these codes in \cite{chen-web}.

\section*{Acknowledgements} The authors would like to thank M. Greferath for his course at Claude Shannon Institute and P. Beelen and T.  H\o holdt for helpful comments on this paper.

\medskip


\begin{thebibliography}{99}

\bibitem{Blackmore-Norton}
Tim Blackmore and Graham~H. Norton.
\newblock Matrix-product codes over {$\Bbb F\sb q$}.
\newblock {\em Appl. Algebra Engrg. Comm. Comput.}, 12(6):477--500, 2001.

\bibitem{ma}
Wieb Bosma, John Cannon, and Catherine Playoust.
\newblock The magma algebra system. {I}. the user language.
\newblock {\em J. Symbolic Comput.}, 24(3-4):235--265, 1997.

\bibitem{chen-web}
Eric~Zhi Chen.
\newblock Web database of binary {QC} codes.
\newblock Online available at
  \url{http://www.tec.hkr.se/~chen/research/codes/searchqc2.htm}.
\newblock Accessed on 2009-03-27.

\bibitem{cota}
Markus Grassl.
\newblock {Bounds on the minimum distance of linear codes}.
\newblock Online available at \url{http://www.codetables.de}, 2007.
\newblock Accessed on 2009-03-27.

\bibitem{hlr}
Fernando Hernando, Kristine Lally, and Diego Ruano.
\newblock Construction and decoding of matrix-product codes from nested codes.
\newblock {\em Appl. Algebra Engrg. Comm. Comput.}, 20:497--507, 2009.

\bibitem{Kas}
T.~Kasami.
\newblock A {G}ilbert-{V}arshamov bound for quasi-cyclic codes of rate {$1/2$}.
\newblock {\em IEEE Trans. Information Theory}, IT-20:679, 1974.

\bibitem{LF}
Kristine Lally and Patrick Fitzpatrick.
\newblock Algebraic structure of quasicyclic codes.
\newblock {\em Discrete Appl. Math.}, 111(1-2):157--175, 2001.

\bibitem{Ozbudak}
Ferruh {\"O}zbudak and Henning Stichtenoth.
\newblock Note on {N}iederreiter-{X}ing's propagation rule for linear codes.
\newblock {\em Appl. Algebra Engrg. Comm. Comput.}, 13(1):53--56, 2002.
\end{thebibliography}
\end{document}